\documentstyle[12pt,aps,epsfig,prl]{revtex}

\begin{document}
\preprint{prl}
\draft

\title{Determination of the Fermi Velocity by Angle-dependent Periodic Orbit Resonance Measurements in the Organic Conductor
$\alpha$-(BEDT-TTF)${_2}$KHg(SCN){$_4$}}

\bigskip

\author{A. E. Kovalev, S. Hill\cite{email}}
\address{Department of Physics, University of Florida, Gainesville, FL 32611}

\author{J. S. Qualls}
\address{Department of Physics, Wake Forest University, Winston-Salem, NC 27109}

\date{\today}
\maketitle

\bigskip

\begin{abstract}

We report detailed angle-dependent studies of the microwave
($\nu=50$ to 90 GHz) interlayer magneto-electrodynamics of a
single crystal sample of the organic charge-density-wave (CDW)
conductor $\alpha$-(BEDT-TTF)${_2}$KHg(SCN){$_4$}. Recently
developed instrumentation enables both magnetic field ({\bf B})
sweeps for a fixed sample orientation and, for the first time,
angle sweeps at fixed $\nu/${\bf B}. We observe series' of
resonant absorptions which we attribute to periodic orbit
resonances (POR) $-$ a phenomenon closely related to cyclotron
resonance. The angle dependence of the POR indicate that they are
associated with the low temperature quasi-one-dimensional (Q1D)
Fermi surface (FS) of the title compound; indeed, all of the
resonance peaks collapse beautifully onto a single set of
$\nu/${\bf B} versus angle curves, generated using a semiclassical
magneto-transport theory for a single Q1D FS. We show that Q1D POR
measurements provide one of the most direct methods for
determining the Fermi velocity, without any detailed assumptions
concerning the bandstructure; our analysis yields an average value
of $v_F=6.5\times10^4$~m/s. Quantitative analysis of the POR
harmonic content indicates that the Q1D FS is strongly corrugated.
This is consistent with the assumption that the low-temperature FS
derives from a reconstruction of the high temperature
quasi-two-dimensional FS, caused by the CDW instability. Detailed
analysis of the angle dependence of the POR yields parameters
associated with the CDW superstructure which are consistent with
published results. Finally, we address the issue as to whether or
not the interlayer electrodynamics are coherent in the title
compound. We obtain a relaxation time from the POR linewidths
which is considerably longer than the interlayer hopping time,
indicating that the transport in this direction {\em is} coherent.

\end{abstract}

\bigskip


\section{Introduction}

\noindent{A detailed knowledge of the Fermi surface (FS)
topologies of low-dimensional conductors is an essential starting
point for understanding the mechanisms that drive the various
electronic instabilities which result in, {\em e.g}. the magnetism
or superconductivity in these systems \cite{ishiguro,singleton1}.
For example, recent theoretical studies have shown that the
symmetry of the superconducting state in quasi-two-dimensional
(Q2D) and quasi-one-dimensional (Q1D) systems is extremely
sensitive to the nesting characteristics of the FS
\cite{tanuma,kuwabara,takimoto}. Furthermore, the organic compound
central to this investigation has recently aroused considerable
interest due to a range of exotic phenomena which derive from its
intrinsic electronic low-dimensionality, {\em e.g.} multiple
field-induced charge density wave (CDW) phases \cite{harrison1},
and field-induced dynamic diamagnetism associated with a
dissipationless conductivity~\cite{harrison2}. Although there
exists an extensive array of experimental techniques for probing
FS topologies, few possess the necessary resolution to profile the
small (often~$<1\%$) deformations (warpings) that arise due to
weak dispersion along the low conductivity axis (axes) of Q2D
(Q1D) systems.}

We have recently developed new methods for determining the FS
topologies of quasi-low-dimensional systems using a
millimeter-wave spectroscopic
technique~\cite{hillpor,mmrsi,lambda}. A novel type of cyclotron
resonance (CR) is predicted to occur $-$ the so-called periodic
orbit resonance (POR) $-$ which is fundamentally different from
the conventional CR observed in normal
metals~\cite{hillpor,blundell2,ardavan1,ardavan2,osadaamro}. This
technique, which was first considered by Osada {\em et al.}
\cite{osadaamro}, essentially corresponds to high frequency
angle-dependent magnetoresistance oscillations
(AMRO)~\cite{singleton1,blundell2,osadaamro}. In this article, we
report extensive high sensitivity POR measurements of the organic
charge transfer salt $\alpha$-(BEDT-TTF)${_2}$KHg(SCN){$_4$}
(where BEDT-TTF denotes bis-ethylenedithio-tetrathiafulvalene or
ET for short~\cite{ishiguro,singleton1}). These studies extend the
earlier work of Ardavan {\em et al.}~\cite{ardavan1}. In
particular, the exceptional sensitivity offered by our
spectrometer, together with the elimination of all spurious
experimental artifacts, enables us to observe many POR harmonics.
The use of a split-pair magnet allows in-situ rotation, thereby
facilitating extensive angle dependent POR investigations.
Consequently, these studies provide a rigorous test of the
semiclassical theories commonly used to analyze many aspects of
the transport properties of low-dimensional conductors
\cite{osadaamro,blundell1,mckenzie1,mckenzie2}, and enable a
determination of important parameters describing the FS topology
of $\alpha$-(ET)${_2}$KHg(SCN){$_4$}. Furthermore, as we shall
show, the POR technique represents one of the most direct and
accurate means of measuring the Fermi velocity in Q1D systems;
here, we report the first such measurement for a Q1D conductor.

$\alpha$-(ET)${_2}$KHg(SCN){$_4$} possesses a layered crystal
structure in which the highly conducting ET planes are separated
by insulating anion layers~\cite{ishiguro,singleton1}; for this
material, the least conducting direction is along the
crystallographic $b^*-$axis. The ratio of in-plane to interlayer
conductivities is about
$\sigma_{\parallel}/\sigma_{\perp}\sim10^5$ (see
Refs.~\cite{ishiguro,singleton1}). Within the conducting layers,
the flat donor ET molecules are arranged face-to-face in a herring
bone structure. Weak overlap of the sulphur $\pi-$orbitals,
oriented normal to the planes of the ET molecules, gives rise to a
fairly isotropic in-plane conductivity, and to a Q2D band
structure which may be calculated using a tight binding
approximation \cite{ishiguro,singleton1,mori}.

According to band structure calculations, the FS of this metal
consists of a pair of Q1D open sections (corrugated sheets), and a
Q2D cylindrical section with the axis of the cylinder
perpendicular to the layers (see Fig.
1a)~\cite{ishiguro,singleton1,mori}. At about 8K,
$\alpha$-(ET)${_2}$KHg(SCN){$_4$} exhibits a phase transition into
a low temperature state \cite{sasakitoyota}; this transition is
believed to be driven by a nesting instability associated with the
Q1D FS, resulting in the formation of a CDW state
\cite{harrison1}. The superstructure associated with the CDW also
affects the Q2D FS~\cite{karts1,harrison3}; the reconstructed
low-temperature FS again consists of both Q1D and Q2D sections, a
possible form of which is shown in Fig. 1b~\cite{karts1}. However,
the origin of the low temperature Q1D FS is not related to the
underlying lattice symmetry; rather, it is related to the CDW
nesting vector. As discussed above, the interplay between the CDW
and the residual low-temperature carriers gives rise to several
spectacular effects in high magnetic fields
\cite{harrison1,harrison2}. Consequently, these investigations
provide important insights into this high field behavior.

The article is organized as follows: in the following section, we
outline the theoretical background behind the POR phenomenon; in
Section III we describe our experimental methods; in Section IV we
present the results of our POR measurements on
$\alpha$-(ET)${_2}$KHg(SCN){$_4$}; the results are discussed in
Section V; and we end with a summary and conclusions in Section
VI.

\section{theoretical background}

\noindent{The energy barrier separating the Q1D and Q2D FS
sections in Fig. 1b is relatively small: in fields above $\sim
10$T, magnetic breakdown occurs, and electrons begin to follow
closed orbits corresponding to the original Q2D FS. These closed
orbits reveal themselves in the Shubnikov-de Haas (SdH) and de
Haas-van Alphen (dHvA) effects~\cite{osadakhg,mass}. The existence
of a highly corrugated Q1D FS was first confirmed by AMRO
measurements in the limit of zero frequency (DC
AMRO)~\cite{osadaamro,blundell1,karts2,iye,caulfield}, {\em i.e.}
under the condition $\nu\ll\tilde{\nu}_c$, where $\nu$ is the
measurement frequency and $\tilde{\nu}_c$ is the frequency with
which electrons cross the Brillouin zone. These measurements
demonstrated that, while rotating magnetic field in a plane
perpendicular to the layers, the AMRO exhibit sharp minima.}

We briefly discuss the physical origin of the Q1D AMRO effect with
reference to the reciprocal space coordinate system defined in
Fig.~2a, in which the Q1D FS is parallel to the $k_yk_z$-plane,
and the highly conducting ET layers are parallel to the
$k_xk_y$-plane \cite{footnote1}. Application of a magnetic field
causes quasiparticles to follow trajectories perpendicular to the
applied field and parallel to the Q1D FS sheets ({\em yz}-plane),
as depicted in Figs. 2b and c. This results in periodic
modulations of the quasiparticle velocities
[$\vec{v}=\hbar^{-1}\nabla_k\varepsilon(k)$, {\em i.e.} $\bot$
FS], which are most pronounced for the components parallel to the
Q1D sheets ($v_y$ and $v_z$ in this case); the resultant
real-space trajectory is illustrated in Fig. 2b. The velocity
modulations are related to the weak FS corrugations, which may be
expressed in terms of Fourier
components~\cite{osadaamro,blundell1}. Each Fourier component has
an associated AMRO minimum. This is best illustrated with the aid
of Figs. 2b and c, and by considering the semiclassical Boltzmann
transport equation \cite{boltzmann}: for general quasiparticle
trajectories on the Q1D FS (Fig. 2b), $v_y$ and $v_z$ are
effectively averaged to zero, and do not contribute appreciably to
the conductivity; for special cases in which quasiparticles follow
trajectories parallel to a particular corrugation (Fig. 2c), $v_y$
and $v_z$ are {\em not} effectively averaged to zero $-$ such
trajectories contribute maximally to the conductivity, thereby
giving rise to the pronounced $\rho_y$ and $\rho_z$ resistivity
minima.

Provided that the FS corrugations are weak, all quasiparticle
trajectories will be parallel and approximately straight in
reciprocal space (not in real space $-$ see Fig. 2b), with their
orientations determined solely by the projection of the applied
magnetic field onto the plane of the FS ({\em yz-}plane)
\cite{osadaamro,blundell1}; AMRO minima then occur whenever the
quasiparticle trajectories run parallel to a particular Fourier
component of the corrugations. Thus, the DC AMRO minimum condition
depends only on the field orientation relative to the
crystallographic (or CDW superstructure) axes, and not on its
magnitude. For an orthorhombic crystal, one expects AMRO minima
which are periodic in $\tan\theta$, and symmetric about
$\theta=0^o$, where $\theta$ is the angle between the applied
field and the normal to the conducting layers [{\em b*}-axis for
$\alpha$-(ET)${_2}$KHg(SCN){$_4$}]. The periodicity is determined
by the crystal (or CDW super-) lattice parameters, and scaled by
the factor $1/\cos\phi$, which projects the applied magnetic field
onto the plane of the FS; $\phi$ is the azimuthal angle between
the {\em yz-}plane, and the plane defined by the magnetic field
and {\em z}-axes (see Fig. 2a). For an oblique lattice, Q1D DC
AMRO minima are observed for $\phi=0$ rotations at angles
$\theta_{mn}$ given by

\begin{equation}
\label{DCAMRO} \tan \theta_{mn}  = \left( {\frac{m}{n} \times
\frac{l}{b^*}} \right) + \frac{l'}{b^*}\\,
\end{equation}

\noindent{where {\em b*} is the interlayer spacing, {\em l} is the
in-plane lattice spacing (parallel to the Q1D FS, or {\em y} in
this case), and $l'$ is the obliquity parameter defined in Fig. 2d
\cite{osadaamro,blundell1}. The indices {\em m} and {\em n}
parameterize the Fourier components of the FS corrugations. For
$\alpha$-(ET)${_2}$KHg(SCN){$_4$}, significant DC AMRO minima have
only been observed for $n=0,1$, and $m=\pm 0,\pm 1,\pm 2,...{\em
etc}.$; $\theta$ rotation measurements, performed at several
different azimuthal angles $\phi$, have enabled determination of
the orientation of the Q1D FS sheet relative to the
crystallographic axes, and of the ratios $l/b^*$ and $l'/b^*$
which characterize the low temperature
CDW~\cite{karts2,iye,caulfield}.}

The theory of both Q1D and Q2D AMRO has been extended to high
frequencies by several
authors~\cite{hillpor,blundell2,ardavan1,ardavan2,osadaamro}; high
frequency implies $\nu\sim\tilde{\nu}_c$, {\em i.e.} when the
measurement frequency becomes comparable to the frequency of the
periodic {\em k-}space orbits. High frequency AMRO, or POR, are
closely related to CR. Indeed, DC AMRO represent a limiting case
of CR/POR. For the situation depicted in Fig. 2c, the
quasiparticle reciprocal space trajectoriy does not modulate the
real space velocity. Consequently, such orbits give rise to a
$\nu=0$ Drude peak in the conductivity, {\em i.e.} an AMRO
minimum~\cite{blundell1}. For the situation depicted in Fig. 2b,
the quasiparticle reciprocal space trajectories {\em do} modulate
the real space velocity components. Such orbits give rise to
finite frequency Drude conductivity peaks. Therefore, although the
situation depicted in Fig. 2b does not give rise to a DC AMRO
minimum, {\em it will} give rise to a finite frequency Drude peak
in the conductivity when the measurement frequency matches the
velocity modulation frequency. In practice, a sample is placed in
a microwave resonator, providing a well defined electromagnetic
field environment, and resonant absorption occurs whenever the
high frequency AMRO/POR condition is met \cite{mmrsi}. As in the
DC case, each Fourier component of warping produces a distinct
POR. However, in contrast to the DC case, the resonance positions
(angles) depend on the ratio of the microwave frequency to the
external field, {\em i.e.} $\nu/B$. The resonance condition for
the orthorhombic lattice, given in
Ref.~\cite{blundell2,osadaamro}, may be generalized for an
arbitrary lattice, and for an arbitrary rotation plane, {\em i.e.}

\

\begin{equation}
\label{rescond}
\frac{\nu}{B_{res}}=\frac{ev_F}{h}\left[(nb^*)^2cos^2\phi+(ml+nl^\prime)^2\right]^{1/2}\left|
sin(\theta-\theta _{mn})\right |,\\
\end{equation}

\noindent{where}

\begin{equation}
\label{zeros} \tan \theta
_{mn}=\frac{1}{cos(\phi)}\left(\frac{ml}{nb^*}+\frac{l^\prime}{b^*}\right),\;\;n=0,1,2...,m=0,\pm
1,\pm 2...
\end{equation}
\

\noindent{Here, {\em b*}, {\em l}, $l'$, {\em m}, {\em n},
$\theta$ and $\phi$ have the same definitions as above; $v_F$ is
the Fermi velocity, which is assumed to be constant over the
entire Q1D FS; and $B_{res}$ represents the applied field strength
at which a POR will be observed for a particular pair of indices
{\em m} and {\em n}. Eq.~(\ref{rescond}) may be further simplified
as follows,}

\

\begin{equation}
\label{rescond2} \frac{\nu}{B_{res}}= \frac{ev_F}{h}\times
nb^*\cos\phi\left| \frac{sin(\theta-\theta _{mn})}{cos(\theta
_{mn})}\right |,\;\; n\geq 0.
\end{equation}

\

\noindent{In contrast to the DC case, one can now observe all
$\theta_{mn}$ resonances by sweeping the magnetic field or
frequency with the sample orientation fixed; our experimental
set-up permits the former (see following section). One can also
carry out angle rotations at fixed $\nu/B$. However, unlike the DC
case, where the AMRO minima are determined solely by
Eq.~(\ref{zeros}) [or Eq.~(\ref{DCAMRO})], the finite frequency
resonances are found from the combined solutions to
Eqs.~(\ref{zeros}) and (\ref{rescond2}). For a particular $\nu/B$,
each DC AMRO minimum splits into two finite frequency branches,
having opposite circular polarizations about the {\em x-}axis:
these result from the positive and negative values of
$\sin(\theta-\theta_{mn})$ in Eq.~(\ref{rescond2}); thus, one
branch moves to higher angles with increasing $\nu/B$, the other
to lower angles. Each branch persists up to a maximum $\nu/B$, at
which $\left(\theta-\theta_{mn}\right)=\pm90^o$. The form of the
angle dependence of the Q1D POR may be seen in Fig.~4 of
Refs.~\cite{blundell2,ardavan1}, and in the results section of
this article (Fig. 5 below).}

In a field swept experiment, several resonances are observed,
corresponding to different Fourier components, or harmonics of the
warping. Unlike the harmonics observed in conventional CR
experiments on Q2D and 3D systems with non-parabolic dispersion
\cite{hillpor,blundell2}, the Q1D POR harmonics are, in general,
not commensurate, {\em i.e.} they are not evenly spaced in
$\nu/B$, as has been noted in previous experiments on Q1D organic
conductors \cite{hillpphmf}. At the DC AMRO minimum angles,
however, the harmonics do become commensurate.

There are a number of fitting parameters in
Eqs.~(\ref{rescond})$-$(\ref{rescond2}): in principle, $\theta$
and $\phi$ should be known, though they may also be determined
from measurement (see below), and through comparison with DC AMRO
data; $b^*$ is known from X-ray data \cite{ishiguro};
$\theta_{mn}$ and, hence, {\em l} and $l'$ may be determined
either from $\theta$ sweeps at different $\phi$ and $v/B$, or
field sweeps at different $\theta$ and $\phi$. The final fitting
parameter, $v_F$, is more-or-less insensitive to the other fit
parameters. Therefore, Q1D POR offers one of the most direct and
accurate means of measuring the Fermi velocity in Q1D systems. We
note that McKenzie and Moses have proposed an alternative theory
to describe Q1D POR in low-dimensional organic conductors
\cite{mckenzie1}, though the resonance condition
[Eqs.~(\ref{rescond})$-$(\ref{rescond2})] turns out to be the
same. However, as we will show below, this theory appears to
inappropriate for the case of $\alpha$-(ET)${_2}$KHg(SCN){$_4$}.

As described above, resonances are expected for the conductivity
components parallel to the Q1D FS. More detailed considerations
show that, for the case of layered organic conductors, it is
preferential to measure the interlayer conductivity $-$ both for
the DC case~\cite{osadaamro}, and for POR
measurements~\cite{hillpor,lambda}. Under such conditions, only
resonances with $n=1$ are detected, and their amplitudes are
proportional to the squares of the Fourier amplitudes ($t^2_{m1}$)
of the FS warping components \cite{blundell2}. The resonance line
shape has a Lorenzian form when observed through conductivity
measurements; and the resonance line width is given by
$\tau^{-1}$, where $\tau$ is the relaxation time.

\section{experimental}

\noindent{The high degree of sensitivity required for single
crystal measurements is achieved using a resonant cavity
perturbation technique in combination with a broad-band
Millimeter-wave Vector Network Analyzer (MVNA~\cite{MVNA})
exhibiting an exceptionally good signal-to-noise ratio
\cite{mmrsi}. The MVNA is a phase sensitive, fully sweepable (8 to
350 GHz), superheterodyne source/detection system. For the
purposes of these measurements, a waveguide probe, optimized to
work in the 40 to $\sim 120$~GHz range, was used to couple
radiation to and from a cylindrical resonator which we operate in
transmission mode; the essential details of this instrumentation
are described in detail in ref. \cite{mmrsi}. In order to enable
rotation of the sample relative to the applied magnetic field, we
utilize split-pair magnet with a 7~T horizontal field and a
vertical access. Smooth rotation of the entire rigid microwave
probe, relative to the fixed field, is achieved via a room
temperature stepper motor mounted at the neck of the magnet dewar;
the stepper motor offers $0.1^o$ angle resolution. The source and
detector (harmonic generator and mixer) are bolted rigidly to the
microwave probe; subsequent connection to and from the MVNA is
achieved via flexible coaxial cables \cite{mmrsi}. In this mode of
operation, one can maintain optimal coupling between the
spectrometer and the cavity containing the sample, {\em whilst}
rotating the probe. As discussed in great detail in
ref.~\cite{mmrsi}, good coupling between the various microwave
elements is essential in order to maintain a high sensitivity and
a low noise level. These factors are the main reason for the
vastly improved quality of the POR data presented in the following
section, relative to similar previous investigations
\cite{ardavan1}. Rotation of the entire rigid sample probe also
ensures that the sample sits in a reproducible electromagnetic
field environment for all measurements made on a particular mode
of the cavity.}

The sample, which was grown by standard techniques
\cite{ishiguro}, has a platelet shape ($\sim$dimensions
$0.8\times0.8\times0.17$~mm$^3)$ and was suspended in the center
of the cylindrical cavity by means of a quartz pillar. The axis of
the cylindrical cavity was aligned with the rotation axis, while
the sample {\em b*}-axis was oriented perpendicular to this axis,
{\em i.e.} parallel to the plane of rotation, thereby ensuring
fixed $\phi$ rotations. The $\theta=\pm 90^o$ orientations may be
determined from fixed field angle sweeps. $\phi$ rotations were
carried out by hand, at room temperature, and recorded using a
digital camera attached to a microscope; calibration of $\phi$ was
achieved through subsequent data analysis. The base frequency of
the cavity used for these measurements was $\nu=53.9$~GHz,
corresponding to the TE011 cylindrical mode with a $Q-$factor of
about $10^4$. Higher frequency measurements were also performed on
higher order modes of the same cavity without the need to warm up
the probe. The use of a cavity enables positioning of a single
crystal sample into a well defined electromagnetic field
environment, {\em i.e.} the orientations of the DC and AC
electromagnetic fields relative to the sample's crystallographic
axes are precisely known. In this way, one can systematically
probe any diagonal component of the conductivity tensor
\cite{lambda}. In this particular investigation, measurements were
restricted to modes which probe only the interlayer
electrodynamics; the precise details as to how we achieve this are
described elsewhere \cite{mmrsi,lambda}. Consequently, the
measured absorption is directly proportional to the interlayer
conductivity $\hat{\sigma}_{b^*}(\omega$,{\bf B},T).

All measurements were carried out at a temperature of about 2.2 K.
The temperature was stabilized using a Quantum Design PPMS
variable flow cryostat. Data were obtained for both field sweeps
at constant angle, and angle sweeps at constant field. The data
presented in the following section are limited to rotation in two
planes corresponding to $\phi$=28$^o$ and $\phi$=66$^o$.

\section{results}

\noindent{Figure 3 shows the typical microwave absorption curves
for field swept measurements at different fixed sample
orientations ranging from $\theta=-20^o$ to $+5^o$ in $5^o$ steps
(the traces have been offset for clarity); $\phi=28^o$ in each
case, and the data were all obtained at $\nu=53.9$~GHz (TE011
mode). The most obvious aspect of the data are the peaks in
absorption corresponding to peaks in the interlayer conductivity.
As we shall now detail, these peaks correspond to PORs. The first
point to note is that there is no simple harmonic relationship
between the peak positions ($B_{res}$) within a given trace, {\em
i.e.} the peaks are incommensurate. Each peak may clearly be
associated with a particular POR branch; in other words, there is
a smooth shift in the positions of the peaks belonging to each
branch, as indicated by the dashed lines. Most notable is the
dominant peak in the $\theta=5^o$ data, which splits into two
distinct branches upon rotation. One branch moves rapidly to
higher fields, while the other branch moves to lower fields with a
weaker angle dependence. This angle dependence is completely
incompatible with a Q2D CR theory, as has been noted previously by
several authors \cite{ardavan1,hillpphmf}. As we shall show below,
all of the resonance peaks collapse beautifully onto a single set
of $\nu/B_{res}$ vs $\theta$ curves generated using a Q1D POR
theory for a single Q1D FS. As discussed in the previous section,
each particular POR branch corresponds to a specific Fourier
component of the warping of the Q1D FS, and the resonances in Fig.
3 have been labeled accordingly. The amplitudes of the peaks are
only weakly angle dependent; we discuss the relative intensities
of the peaks further below. The quality of the data in Fig.~3 is
unprecedented for a measurement in this frequency range; indeed,
this data is of comparable quality to the best DC measurements.}

Figure~4 shows microwave absorption for fixed $\nu/B$
$\theta$-sweeps, at an azimuthal angle $\phi$=28$^o$. These data
were again obtained using the TE011 mode of the cavity, while
varying $\nu/B$ by means of the applied field strength (indicated
in the figure). Once again, several clear absorption peaks are
observed corresponding to distinct PORs. In fact, each trace is
highly reminiscent of DC AMRO data
\cite{osadaamro,blundell1,karts2,iye,caulfield}, albeit inverted,
{\em i.e.} conductivity peaks are seen in POR measurements,
whereas resistivity minima are observed in AMRO experiments.
However, closer inspection indicates that the POR peak positions
clearly depend on the field strength, in marked contrast to the DC
AMRO case. Once again, each peak belongs to a POR branch
corresponding to a particular Fourier component of the FS warping,
and several of the branches have been labeled accordingly in
Fig.~4. The noise level is slightly higher in Fig.~4, when
compared to Fig.~3. This is attributable to unavoidable mechanical
vibrations associated with the rotation of the probe. Thus, the
field swept measurements are intrinsically cleaner. However, the
signal-to-noise ratio is still exceptionally good, even in the
angle swept mode. It is also somewhat easier to observe higher
order (higher $\left|m\right|$) POR in this mode of operation, by
sweeping through angles close to $\theta=\pm 90^o$ at the maximum
field of $\sim7$T. At lower fields, the condition
$2\pi\tilde{\nu}\tau=\tilde{\omega}\tau\sim 1$ is not met for many
of the higher $m$ modes, as evidenced by the disappearance of the
closely spaced peaks around $\theta=90^o$ in the lower field data
in Fig.~4.

In Fig.~5, we compile all of the $\phi$=28$^o$ field swept data
into a single plot of $\nu/B_{res}$ versus $\theta$, where the
data points represent the peak positions, $B_{res}$, obtained from
data similar to those in Fig.~3. Fig.~5 includes measurements
performed at two separate frequencies: $\nu=53.9$~GHz (open
circles) and $\nu=89.2$~GHz (solid triangles). The data collapse
onto a set of $\sin(\theta-\theta_{mn})$ arches according to
Eqs.~(\ref{rescond}) and (\ref{rescond2}). All of the resonances
may be labeled using a single index {\em m}, as is also the case
for DC AMRO measurements~\cite{blundell1,karts2,iye,caulfield}.
The dotted lines are fits according the Eqs.~(\ref{zeros}) and
(\ref{rescond2}); the exceptional quality of the fits may be
attributed to the high quality of the measurements. Qualitative
comparisons of the intensities of the peaks reveal that they scale
approximately with $\cos\theta_{mn}$, {\em i.e.} the resonances
with zero field intercepts closest to $\theta=0^o$ ($m=0$ and $1$
Fig. 5) have the strongest intensities.

In order to compare data obtained for different rotation planes
(different $\phi$), we expand the $\sin(\theta-\theta_{mn})$ term
in Eq.~(\ref{rescond2}), giving

\

\begin{equation}
\label{rescond3} \frac{\nu}{B_{res}\cos\theta}=
\frac{ev_F}{h}\times
b^*\cos\phi\left|\tan\theta_{mn}-\tan\theta\right |,
\end{equation}

\

\noindent{where we have set $n=1$. Thus, plots of
$\nu/(B_{res}\cos\theta)$ versus $\tan\theta\cos\phi$ should
produce straight lines of slope $\pm eb^*v_F/h$, with offsets
given by $\theta_{mn}$. Such a plot is shown in Fig.~6, for all of
the data obtained from these investigations, {\em i.e.} both field
and angle sweeps, and for two rotation planes ($\phi=28^o$ and
$\phi=66^o$). We see that the data collapse very nicely onto the
theory over most of the angle range. Deviations from the theory
may be mainly attributed to small errors associated with the
calibration of the true angle $\theta$; the nature of the theory
amplifies these errors for angles close to $\theta=90^o$, {\em
i.e.} data close to the edges of Fig.~6.}

From the fits in Figs.~5 and~6, we can extract a Fermi velocity
for the Q1D Fermi surface of about $v_F=6.5\times 10^4 m/s$. In
principle, the same value could be estimated from the data in
Ref.~\cite{ardavan1}. This value for the Fermi velocity is very
close to the value obtained from fits to optical conductivity data
using a Drude model~\cite{tamura}. It should be stressed, however,
that our measurements give the most straightforward determination
of the Fermi velocity; the only assumption that is made is that
the PORs are due to the semiclassical motion of quasiparticles on
a Q1D FS $-$ no assumptions about the details of the band
structure are required! Nevertheless, this represents an average
value, {\em i.e.} we have assumed a constant Fermi velocity over
the entire FS. We shall comment on the validity of this assumption
in the following section.

The ratios $l/b^*$ and $l^\prime/b^*$ obtained from our data are
1.2 and 0.5 respectively. These values are in excellent agreement
with those obtained from DC AMRO \cite{karts2,iye,caulfield}:
$l^\prime/b^* \approx$0.5, $l/b^*=1.25-1.35$. Using the
theoretical prediction for the line
shape~\cite{osadaamro,blundell2}, we can estimate the relaxation
time $\tau=1.5\times$10$^{-11}$~s, which is approximately the same
for all peaks. This value is rather high in comparison to the
typical $\tau$ values derived from a Dingle analysis of SdH and
dHvA oscillations. However, similar discrepancies have been noted
in the past from POR data obtained for other Q2D organic
conductors \cite{hillpor}. The differences may be attributed to
sample inhmogeneities, which give rise to additional damping of
SdH/dHvA oscillations. This damping is incorporated into the
theory of the SdH and dHvA effects as an effective scattering time
\cite{shoenberg}. Consequently, Dingle analysis likely under
estimates $\tau$ significantly.

\section{discussion}

\noindent{Comparisons between our data and those obtained by
Ardavan {\em et al}.~\cite{ardavan1}, reveal considerable
advantages to our technique. The rotating cavity design utilized
by Ardavan {\em et al}. offers the main benefit that it may be
used in axial high field magnets providing fields of up to
45~tesla at the National High Magnetic Field Laboratory (NHMFL) in
Florida \cite{splitpair}. However, the split-pair configuration
offers many other important advantages that outweigh the high
field capability. In particular, the use of a rigidly coupled
probe results in a marked improvement in sensitivity ($Q-$factors
of up to 25,000), and a significant reduction in noise
\cite{mmrsi}. Furthermore, data acquisition is possible whilst
simultaneously rotating the field. Gold plating of the waveguides
at the high field end of the probe also eliminates spurious
magnetic resonances associated with contaminants \cite{mmrsi};
these are dramatically apparent in the data in
Ref.~\cite{ardavan1}, thereby obscuring important details of the
POR measurements. All of these factors add up to a marked
improvement in the quality of the results presented here, in
comparison to previously published measurements on
$\alpha$-(ET)${_2}$KHg(SCN){$_4$}. This has enabled us to carry
out the most detailed magneto-optical investigation of the complex
low-temperature phase of this material to date.}

The observation of a strong harmonic content to the POR implies
that the FS is strongly corrugated, thereby suggesting comparable
transfer integrals between neighboring and more distant molecules.
This conclusion contradicts the fact that band structure is well
described by a tight binding approximation. However, as pointed
out in the introduction, the low temperature Q1D FS in
$\alpha$-(ET)${_2}$KHg(SCN){$_4$} results from a reconstruction of
the high temperature FS obtained from tight-binding
calculations~\cite{karts1,harrison3}. The reconstruction is caused
by the fairly weak superstructure associated with a low
temperature CDW transition, {\em i.e.} the Q1D FS is essentially
pieced together from sections of the original high temperature Q2D
closed FS. Thus, the transfer integrals that one could deduce from
the POR data do not characterize the tight binding nature of the
FS. Rather, they characterize the CDW nesting vector and the
original Q2D FS. Indeed, the Fermi velocity obtained from these
measurements corresponds precisely to the Q2D Fermi velocity in
the high temperature state.

We mentioned above, that the POR peak amplitudes are about the
same for resonances having the same
$\left|\cos\theta_{mn}\right|$. This implies that the POR peak
amplitudes are closely related to the FS corrugation wavevectors
$Q\sim b^*/\left|\cos\theta_{mn}\right|$. Comparisons between the
$m=0,1$ POR peaks, with the $m=2,-1$ peaks, reveals a roughly
factor of 6 difference in amplitude, while the difference in
$\left|\cos\theta_{mn}\right|$ is about 2 for these resonances. As
pointed out in Section~II, the POR amplitudes should scale as the
squared Fourier amplitudes associated with the FS corrugation. We
note that, for the most extreme form of warping (a square wave),
the POR amplitudes would scale as $1/Q^2$. Consequently, the $1/6$
difference in amplitude between PORs corresponding to FS warping
components differing in $Q$ by a factor of 2, indicates that the
corrugation is close to this extreme limit. Thus, it is possible
that a reconstruction along the lines discussed in
Ref.~\cite{harrison3} offers the most realistic description for
the low temperature FS in $\alpha$-(ET)${_2}$KHg(SCN){$_4$}. Due
to the strong corrugation, it is quite likely that our assumption
of a constant Fermi velocity over the Q1D FS breaks down to some
extent. Furthermore, the theory for the POR developed in
section~II is only approximate when the FS warping is strong.
However, deviations from this theory are only likely to be
significant when the magnetic field is tilted appreciably out of
the plane of the Q1D FS (see Ref.~\cite{blundell1}), which may
offer an alternative explanation for the deviations between theory
and experiment at the edges of Fig.~6.

McKenzie and Moses have published an alternative theory for
POR~\cite{mckenzie1}. However, this theory predicts strong
so-called "Danner oscillations" (see
refs.~\cite{mckenzie2,danner}) for field rotations close to
$\theta=90^o$ in the $\phi=90^o$ plane, {\em i.e.} the rotation
plane perpendicular to the Q1D FS; these oscillations have never
been observed in $\alpha$-(ET)${_2}$KHg(SCN){$_4$}. This theory
also predicts a strong angle dependence of the POR amplitudes for
all but the $m=0$ harmonic, which also contradicts our results.
However, McKenzie and Moses raise the important question as to
whether interlayer transport should be coherent in order to
observe POR and AMRO \cite{mckenzie2}. This question has since
been re-addressed by numerous authors (see {\em
e.g.}~\cite{singleton3}). We believe that coherent interlayer
transport is not a necessary condition. Within the tight binding
approximation, there should be no difference whether one considers
"hopping" between layers, or coherent motion; in both of the
cases, only the orbital overlaps between nearest neighbor
molecules is important.

If one assumes that the interlayer transport is "weakly
incoherent" (according to the definition of McKenzie
\cite{mckenzie2}), then energy and in-plane momentum should be
conserved for interlayer hopping. For the sake of simplicity, we
consider this situation for the case of a magnetic field rotated
in the {\em zy}-plane, {\em i.e.} parallel to the Q1D FS. In the
presence of both the DC magnetic field, and an electromagnetic
field, the conservation equations are:

\
\begin{eqnarray}
\label{energy} E^f - E^{i}= h\nu = \hbar v_F(k_x^{f}-k_x^{i})\cdot
sign(k_x)\
\end{eqnarray}

\noindent{and}

\begin{eqnarray}
\label{momentum} \hbar {\bf k}_\| ^{i}+ e{\bf A}_\|^i = \hbar {\bf
k}_\|^{f} + e{\bf A}_\|^{f},
\end{eqnarray}
\

\noindent{where {\em E} is the quasiparticle energy; $\hbar{\bf
k}_\|$ is the quasiparticle momentum parallel to the layers; ${\bf
A}_\|$ is the component of the magnetic vector potential parallel
to the layers at the site of the hopping; and the superscripts $i$
and $f$ denote the initial and final energy, momentum, {\em etc.}.
Because the magnetic field only has $B_z$ and $B_y$ components,
one may choose a gauge in which $A_x$ is the only non-zero
component of the vector potential, {\em i.e.} $A_x=B_y z-B_z y$.
The change in the vector potential for interlayer hopping is then
$A^{f}_x-A^i_x=B_y b^*-B_z(ml+l^\prime)$, leading to}

\
\begin{equation}
\label{hop} \nu=v_F\frac{e}{h}|A_x ^{f} - A_x^i|=\frac{e v_F B
}{h}\left|b^* \sin\theta-(ml+l^\prime) \cos\theta\right|
\end{equation}
\

\noindent{which is equivalent to the Eq.~(\ref{rescond3}).
Consequently, the only way to check whether the interlayer
transport is coherent is to compare the interlayer hopping rate
$\tau_h$ $(=\hbar/t_\perp)$, with the relaxation time $\tau$
obtained from the POR linewidths. An estimation of the interlayer
transfer integral from the known conductivity anisotropy gives
$t_\perp \sim 1$~meV, leading to $\tau/\tau_h\sim$~20, indicating
that the interlayer transport {\em is} coherent. The factor of 10
or so difference between the scattering times deduced from
magneto-optical studies, and from SdH or dHvA measurements, raises
some important issues. We believe that the scattering times
deduced from SdH and dHvA measurements can be misleading, since
they do not necessarily reflect the true short-range inelastic
scattering time. As discussed above, this may be caused by the
effects of sample inhomogeneities on the SdH/dHvA oscillation
amplitudes, which are indistinguishable from the effects of real
scattering processes~\cite{shoenberg}.

\section{summary and conclusions}

\noindent{We have presented detailed angle-dependent POR
measurements for the organic CDW conductor
$\alpha$-(ET)${_2}$KHg(SCN){$_4$}. In particular, we demonstrate
the huge potential of angle swept POR measurements. The quality of
our data is unprecedented for measurements in this high frequency
range; indeed, it is of comparable quality to the best DC
measurements. Extensive angle dependent measurements confirm that
the POR are associated with the low temperature Q1D FS; indeed,
all of the resonance peaks collapse beautifully onto a single set
of $\nu/${\bf B} versus angle curves, generated using a
semiclassical magneto-transport theory for a single Q1D FS.
Quantitative analysis of the POR harmonic content indicates that
the Q1D FS is strongly corrugated, a fact that is consistent with
the assumption that the low-temperature FS derives from a
reconstruction of the high temperature quasi-two-dimensional FS.
Extrapolations of our data to zero frequency reveal good agreement
with published DC AMRO measurements.}

We argue that Q1D POR measurements provide one of the most direct
methods for determining the Fermi velocity in Q1D systems $-$ our
analysis yields an average value of $v_F=6.5\times10^4$~m/s.
Furthermore, we show that POR is possible both for coherent and
incoherent interlayer transport. However, based on the
Mott-Ioffe-Regel, the interlayer transport appears to be coherent
in this compound, {\em i.e.} $\tau/\tau_h\sim$20.

\section{Acknowledgements}

\bigskip

\noindent{We thank S. Khan and N. Bushong for assistance. This
work was supported by the National Science Foundation (DMR0196461
and DMR0196430). S. H. would like to thank the Research
Corporation for financial support.}

\clearpage

\noindent{\bf Figure captions}

\

\noindent{Fig. 1a) Room temperature FS of
$\alpha$-(ET)${_2}$KHg(SCN){$_4$} according to the calculation in
Ref.~\cite{mori}. Below 8~K, this FS undergoes a reconstruction,
as shown in b); the reconstruction is caused by the
superstructure, with characteristic wavevector $Q$, associated
with the low-temperature CDW \cite{karts1}.}

\

\noindent{Fig. 2a) Definition of the angles $\theta$ and $\phi$,
defining the magnetic field ({\bf B}) orientation relative to the
Q1D FS within the reciprocal space coordinate system discussed in
the main text; the Q1D FS is parallel to the $k_yk_z$-plane, and
the normal to the highly conducting layers ($b^*$-axis) is
parallel to $k_z$. b) A schematic of a typical quasiparticle
trajectory in reciprocal space for a magnetic field applied at an
arbitrary angle $\theta$ away from the $k_z$ axis, within the
plane of the Q1D FS ($k_yk_z$-plane); the motion of the
quasiparticle across the FS corrugations results in a periodic
modulation of the real space velocity, $v_R$ (see text for
discussion). c) At certain applied field orientations, the
quasiparticle trajectories follow constant real-space-velocity
contours; these angles correspond to the $\theta_{mn}$ AMRO minima
(see text). d) The oblique real-space crystal lattice, defining
the CDW superstructure within the plane parallel to the Q1D FS;
$b^*$ is the interlayer spacing, {\em l} and $l'$ are defined in
the text.}

\

\noindent{Fig. 3. Microwave absorption curves for field swept
measurements at different fixed sample orientations (indicated in
the figure); the temperature is 2.2~K, and the traces have been
offset for clarity. The POR have been labeled with the appropriate
value of the index {\em m}.}

\

\noindent{Fig. 4. Microwave absorption for fixed $\nu/B$
$\theta$-sweeps (offset for clarity), at an azimuthal angle
$\phi$=28$^o$. $\nu/B$ was varied by means of the applied field
strength (indicated in the figure); the measurement frequency was
$53.9$~GHz and the temperature was 2.2~K. The POR have been
labeled with the appropriate value of the index {\em m}.}

\

\noindent{Fig. 5. A compilation of the angle dependence of all of
the $\phi$=28$^o$ field swept data, where the data points
represent the peak positions, $B_{res}$, obtained from plots
similar to Fig.~3; measurements at two frequencies are included in
the figure. Several of the POR branches have been labeled with the
appropriate value of the index {\em m}.}

\

\noindent{Fig. 6. A compilation of all of the data obtained in
this investigation, scaled according to Eq.~(\ref{rescond3}).
Several of the POR branches have been labeled with the appropriate
value of the index {\em m}.}

\clearpage

\begin{figure}
\centerline{\epsfig{figure=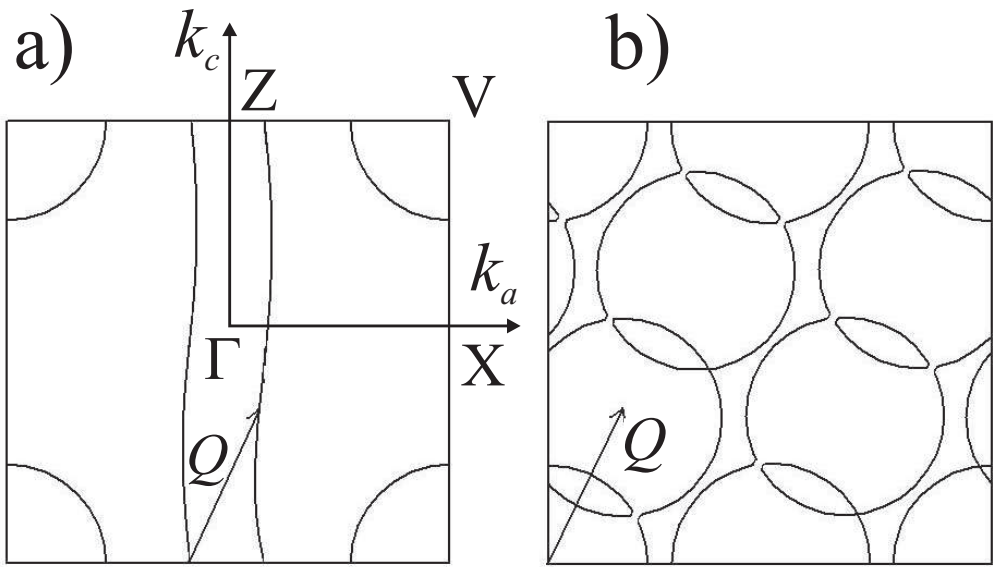,width=140mm}}
\bigskip
\caption{Kovalev {\em et al.}} \label{Fig. 1}
\end{figure}

\clearpage

\begin{figure}
\centerline{\epsfig{figure=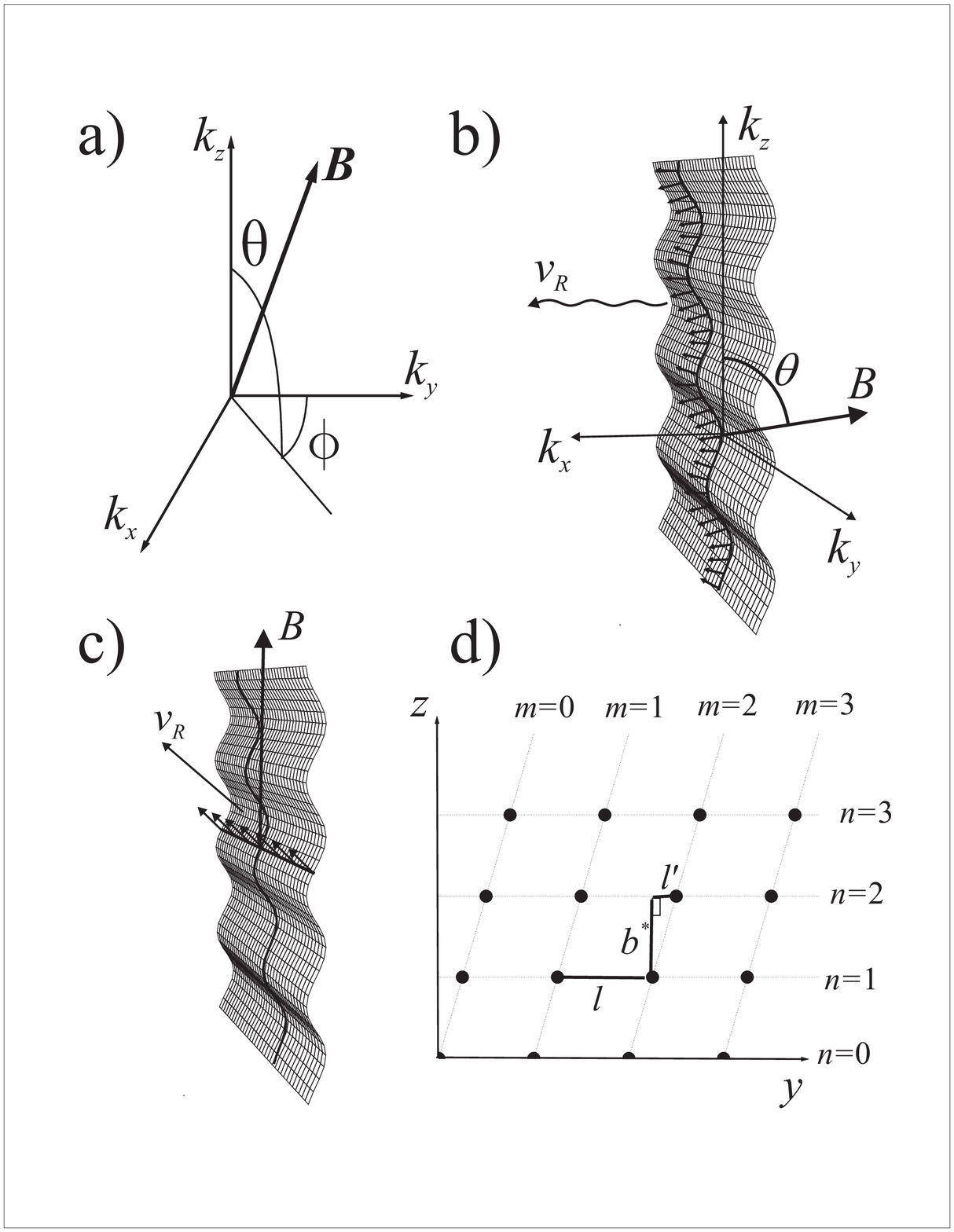,width=140mm}}
\bigskip
\caption{Kovalev {\em et al.}} \label{Fig. 2}
\end{figure}

\clearpage

\begin{figure}
\centerline{\epsfig{figure=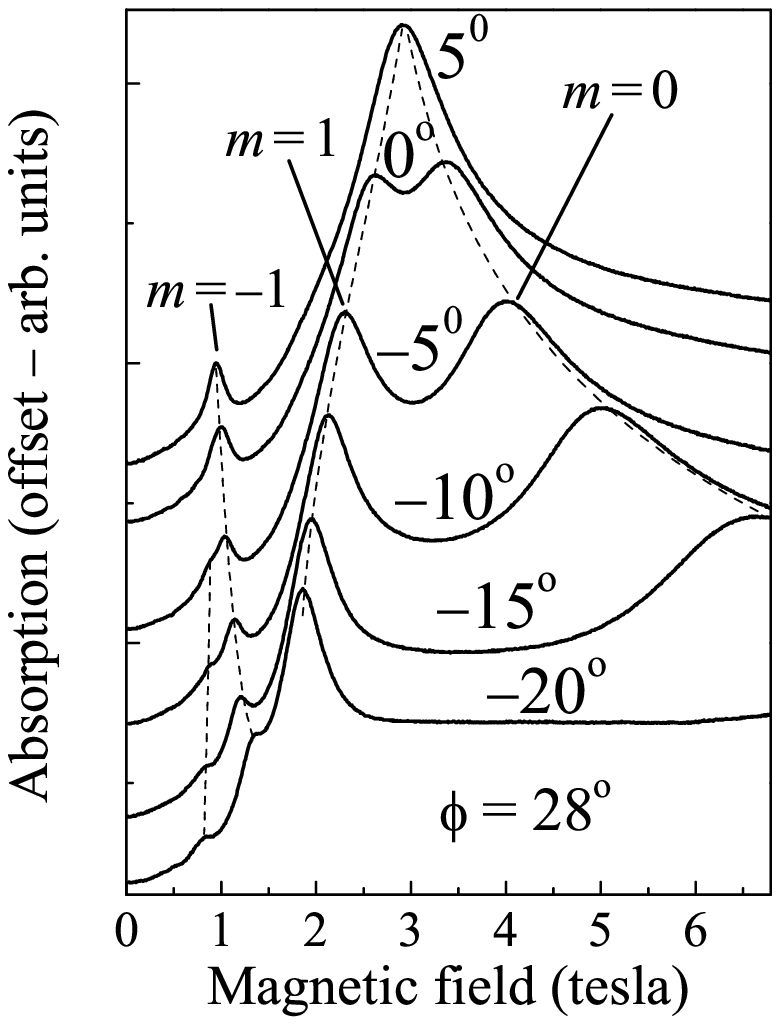,width=120mm}}
\bigskip
\caption{Kovalev {\em et al.}} \label{Fig. 3}
\end{figure}

\clearpage

\begin{figure}
\centerline{\epsfig{figure=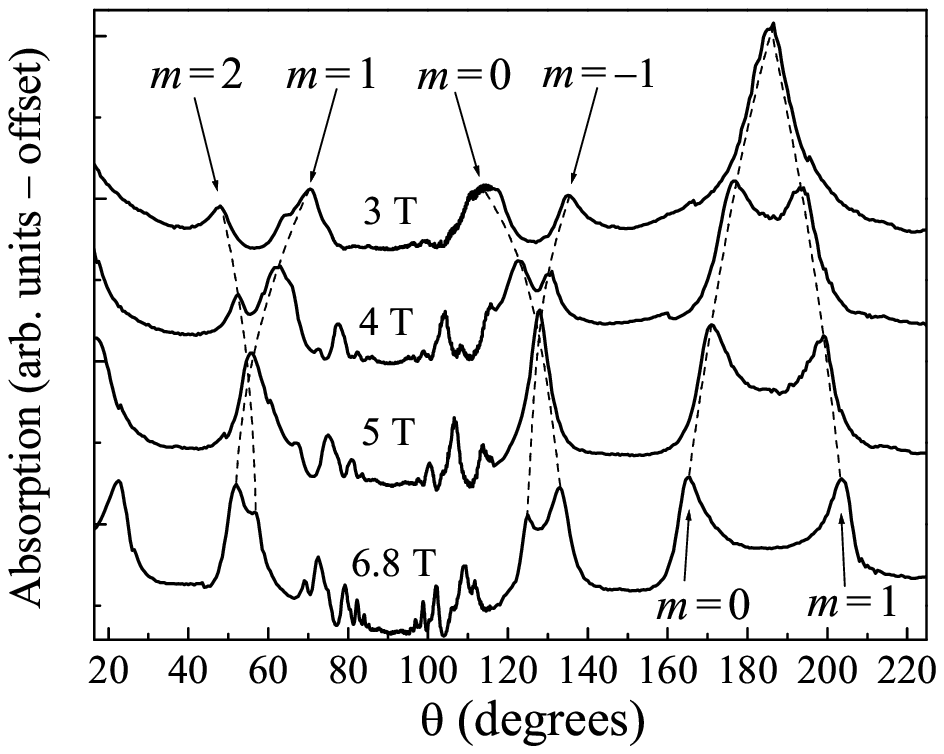,width=180mm}}
\bigskip
\caption{Kovalev {\em et al.}} \label{Fig. 4}
\end{figure}

\clearpage

\begin{figure}
\centerline{\epsfig{figure=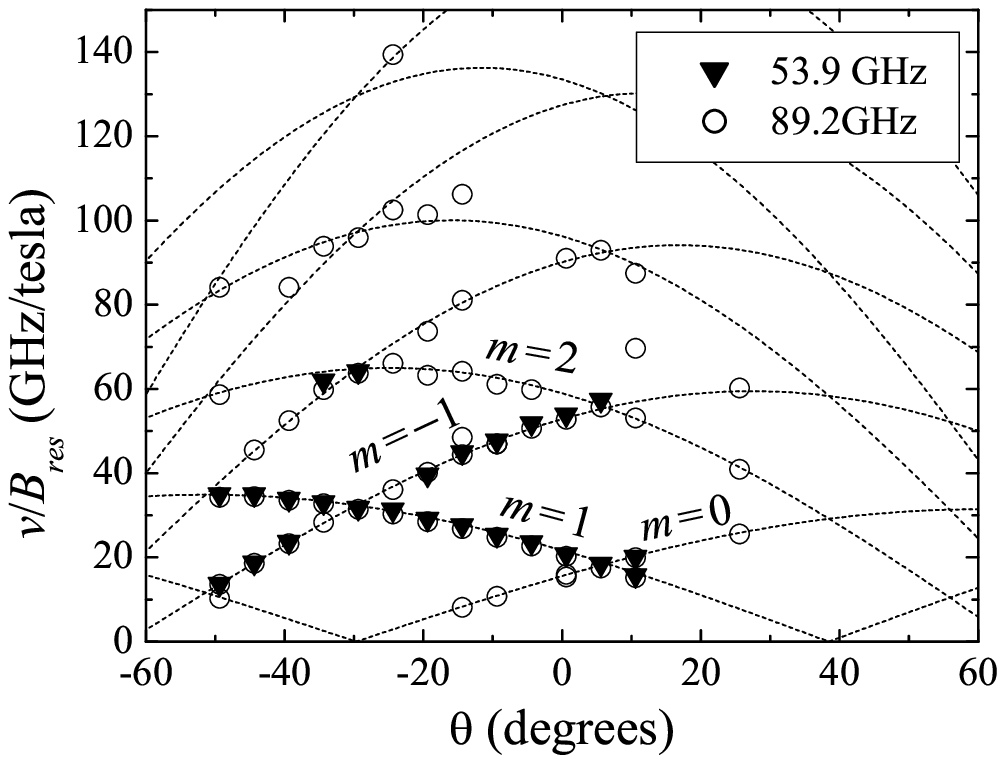,width=180mm}}
\bigskip
\caption{Kovalev {\em et al.}} \label{Fig. 5}
\end{figure}

\clearpage

\begin{figure}
\centerline{\epsfig{figure=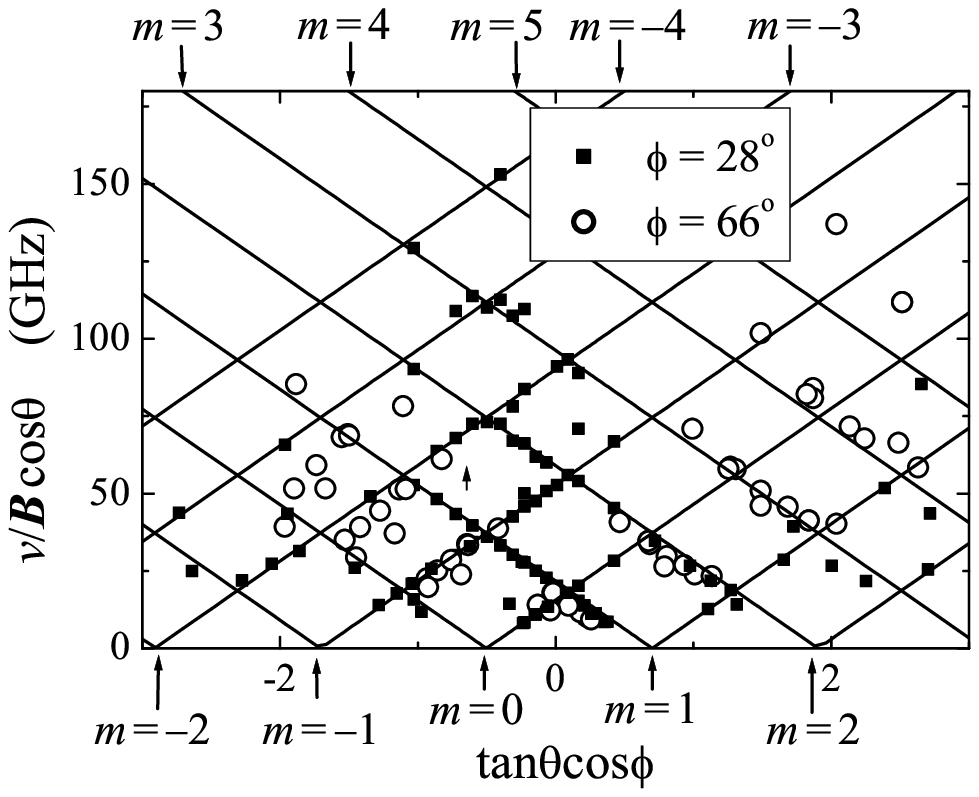,width=180mm}}
\bigskip
\caption{Kovalev {\em et al.}} \label{Fig. 6}
\end{figure}


\begin{references}
\bibitem{splitpair}
We note that a high field split-pair magnet is currently under
construction at the NHMFL, which will enable extension of these
methods to much higher fields in future.
\end{references}
\end{document}